# Three Important Theorems for Fluid Dynamics

Hua-Shu Dou

Temasek Laboratories
National University of Singapore
Singapore 117508

Email: tsldh@nus.edu.sg; huashudou@yahoo.com

**Abstract**: The new proposed "energy gradient theory," which physically explains the phenomena of flow instability and turbulent transition in shear flows and has been shown to be valid for parallel flows, is extended to curved flows in this study. Then, three important theorems for fluid dynamics are deduced. These theorems are (1) Potential flow (inviscid and irrotational) is stable. (2) Inviscid rotational (nonzero vorticity) flow is unstable. (3) Velocity profile with an inflectional point is unstable when there is no work input or output to the system, for both inviscid and viscous flows. These theorems are, for the first time, deduced, and are of great significance for the understanding of generation of turbulence and the explanation of complex flows. From these results, it is concluded that the classical Rayleigh theorem (1880) on inflectional velocity instability of inviscid flows is incorrect which has last for more than a century. It is demonstrated that existence of inflection point on velocity profile is a sufficient condition, but not a necessary condition for flow instability, for both inviscid and viscous flows. In addition, the paradox of dual role of viscosity has been resolved. In parallel flows and Taylor-Couette flows, viscosity has only stable role to the flow owing to its leading to high energy loss which damps instability and there is no viscous diffusion caused by transversal velocity. In non-parallel flows, such as boundary layer flow, viscosity may have unstable role to the flow owing to that viscous diffusion of transversal velocity may change the transversal energy gradient, besides its stable role due to energy loss.

# 1. Introduction

The classical linear stability theory which is back to Rayleigh (1880) has been well established in the literature and in the most text books [1-6]. However, for the onset of instability, this theory obtains agreement with experiments only for few flows cases (Rayleigh-Bernard problem and Taylor-Couette problem), and disagrees with most other flow cases (pipe flow, channel flow, Couette flow, and boundary layer flow). For Poiseuille flow in a straight pipe and plane Couette flow, linear stability analysis shows that they are stable for all the range of Reynolds number while they both transit to turbulence at finite Reynolds number in experiments [1-5]. These phenomena are still not clarified so far. In order to interpreter these discrepancies with the experiments, people even used different explanations for different flows [2]. For example, dual roles of viscosity was assumed in order to explain the turbulent generation in boundary layers [2,7] to merge the disagreement of the theory with experiments. The flow instabilities are artificially divided into viscous instability and inviscid instability [2,3]. Even if these assumptions are made, there are still many flow phenomena could not be well understood.

Energy method based on Reynolds-Orr equation searches for the minimum Reynolds number of the disturbance energy monotonically decrease with the time in the system [1-3,8]. The predicted critical Reynolds number is much lower than that from experiments for parallel flows [2-3]. On the other hand, the occurrence of stability is strictly a local behaviour and the flow during the transition is intermittent. The first occurrence of the flow instability generally takes place in the most "dangerous" positions as seen in the formation of turbulence spot, the cylinder wake, and the dynamic stall on the airfoil with large attack angle. Hence, a method considering the local flow behaviour may be the correct approach.

The weekly nonlinear method which has been developed for a half century [9] and the secondary instability theory which was developed in 1980's [10] seem to give better results than the above-mentioned methods and can explain some phenomena; however there is still discrepancy with experiments.

In a recent study, Dou developed an "Energy Gradient Theory" by rigorous derivation, in which the detail of amplification or decay of the disturbance has been described [11-13]. The theory proposes that in shear flows it is the transverse energy gradient interacting with a disturbance to lead to the flow instability, while the energy loss, due to viscous friction along the streamline, damps the disturbance. The mechanisms of velocity inflection and formation and lift of the hairpin vortex are well explained with the analytical result; the disturbed particle exchanges energy with other particles in transverse direction during the cycle and causes the particle leaves its equilibrium position. The threshold amplitude of disturbance for transition to turbulence is scaled with Re by an exponent of $\gamma = -1$ in parallel flows, which explains the recent experimental result of pipe flow by Hof et al. [14]. The study also confirms the results from asymptotic analysis (for $\text{Re} \to \infty$) of the Navier-Stokes equations by Chapman [15]. He proposed a function of energy gradient and then took the maximum of this function in the flow field, $K_{max}$, as the criterion for flow instability. This approach obtains a consistent value of $K_{max}$ for the critical condition (i.e., minimum Reynolds number) of turbulent transition in parallel flows including plane Poiseuille flow, pipe Poiseuille flow and plane Couette flow [11-13]. In energy gradient theory, the viscosity only plays a stability role because large viscosity could produce large energy loss and thus stabilize the flow. This is in agreement with the experimental observations.

In this paper, the newly proposed energy gradient theory is extended to curved flows. Then, based on the results, three important theorems for fluid dynamics are deduced. From these results, it is concluded that the classical Rayleigh theorem on inflectional velocity instability of inviscid flows is incorrect which has last for more than a century.

## 2. Energy Gradient Theory applied to Curved flows

The energy gradient theory has been described for parallel flows in detail in [11]. Extending the theory from parallel flow to curved flow, we only need to change the Cartesian coordinates (x, y) to curvilinear coordinates (s, n), to change the kinetic energy ($\frac{1}{2}mu^2$) to the total mechanical energy ($E = p + \frac{1}{2}\rho u^2$), and to make the velocity ($u$) along the streamline. Here, we use the same derivation steps as in [11].

Let us consider the *elastic* collision of particles when a disturbance is imposed to the base of a curved shear flow (Fig.1). A fluid particle $P$ at its equilibrium position will move a cycle in vertical direction under a vertical disturbance, and it will have two collisions with two particles ($P_1$ and $P_2$) at its maximum disturbance distances, respectively. The masses of the three particles are $m$, $m_1$ and $m_2$, and the corresponding velocities prior to collisions are $u$, $u_1$ and $u_2$. We use primes for the corresponding quantities after collision. Without lose of generality, we may assume $m = m_1 = m_2$ for convenience of the derivation. For parallel flows, only kinetic energy difference exists between neighboring streamlines. For curved flows, the difference of energy between streamlines is the difference of the total mechanical energy. When fluid particles exchange energy by collisions, it is the exchange of the total mechanical energy. For a cycle of disturbances, the fluid particle may absorb energy by collision in the first half-period and it may release energy in the second half-period because of the gradient of the total mechanical energy. The total momentum and total mechanical energy are conserved during the elastic collisions. The conservation equations for the first collision on streamline $S_1$ are

$$m_1 u_1 + mu = m_1 u'_1 + mu' = \alpha_1 (m_1 + m) u_1, \qquad (1)$$

and

$$Q_1 E_1 + QE = Q_1 E'_1 + QE' = \beta_1 (Q_1 + Q) E_1. \qquad (2)$$

Here $Q = m/\rho$ and $Q_1 = m_1/\rho$ are the volumes of the particles, and $\alpha_1$ and $\beta_1$ are two constants and $\alpha_1 \leq 1$ and $\beta_1 \leq 1$. The values of $\alpha_1$ and $\beta_1$ are related to the residence time of the particle at $P_1$. If the residence time at position $P_1$ is sufficiently long (e.g. whole half-period of disturbance), the particle $P$ would have undergone a large number of collisions with other particles on this streamline and would have the same momentum and total mechanical energy as those on the line of $S_1$, and it is required that $\alpha_1 = 1$ and $\beta_1 = 1$. In this case, the energy gained by the particle $P$ in the half-period is $(QE_1 - QE)$. When the particle $P$ remains on $S_1$ for less than the necessary half-period, of the disturbance, the energy gained by the particle $P$ can be written as $\beta^*_1 (QE_1 - QE)$, where $\beta^*_1$ is a factor of fraction of a half-period with $\beta^*_1 < 1$.

The requirements of conservation of momentum and energy should also be applied for the second collision on streamline $S_2$, and similar equations to Eq.(1) and (2) can be obtained as in [11]. The difference is that the energy gained in the second half-cycle is negative owing to the disturbance of energy in base flow. For the first half-period, the particle gains energy by the collision and the particle also releases energy by collision in the second half-period.

We use the (s, n) to express the coordinates in streamwise and transverse directions, respectively. Using the similar derivations to those in [11], the energy variation of per unit volume of fluid for a half-period for the disturbed fluid particles can be obtained as,

$$\Delta E = \frac{2}{T} \int_0^{T/2} \frac{\partial E}{\partial n} n \, dt = \frac{\partial E}{\partial n} \frac{2}{T} \int_0^{T/2} n \, dt \qquad (3)$$

Where $E = p + (1/2)\rho u^2$ is the total mechanical energy per unit volume of fluid, and T is the period.

Without lose of generality, assuming that the disturbance variation is associated with a sinusoidal function,

$$n = A \sin(\omega t + \varphi_0) \qquad (4)$$

where *A* is the amplitude of disturbance in transverse direction, $\omega$ is the frequency of the disturbance, *t* is the time, and $\varphi_0$ is the initial phase angle. For curved flow, A is respectively expressed by $A_1$ and $A_2$ in the first half and the second half circle, generally, $A_1 \neq A_2$. The velocity of the disturbance in the vertical direction, is the derivative of (4) with respect to time,

$$v' = \frac{dn}{dt} = v'_m \cos(\omega t + \varphi_0). \tag{5}$$

Here, $v'_m = A\omega$ is the amplitude of disturbance velocity and the disturbance has a period of $T = 2\pi/\omega$.

Substituting Eq.(4) into Eq. (3), we obtain the energy variation of per unit volume of fluid for the first half-period,

$$\Delta E = \frac{\partial E}{\partial n} \frac{2}{T} \int_0^{T/2} n dt = \frac{\partial E}{\partial n} \frac{2}{T} \int_0^{T/2} A \sin(\omega t + \varphi_0) dt$$
$$= \frac{\partial E}{\partial n} \frac{2}{T} \frac{1}{\omega} \int_0^{\pi} A \sin(\omega t + \varphi_0) d\omega t = \frac{\partial E}{\partial n} \frac{2A}{\pi}. \tag{6}$$

The stability of the particle can be related to the energy gained by the particle through vertical disturbance and the energy loss due to viscosity along streamline in a half-period.

The energy loss per unit volume of fluid along the streamline due to viscosity in a half-period,

$$\Delta H = \frac{\partial H}{\partial s} l = \frac{\partial H}{\partial s} \frac{\pi}{\omega} u. \tag{7}$$

where *H* is the energy loss per unit volume of fluid due to viscosity along the streamline, $l = u(T/2) = u(\pi/\omega)$ is streamwise length moved by the particle in a half-period.

The magnitudes of $\Delta E$ and $\Delta H$ determine the stability of the flow. After the particle moves a half cycle, if the net energy gained by collisions is zero, this particle will stay in its original equilibrium position (streamline). If the net energy gained by collisions is larger than zero, this particle will be able to move into equilibrium with a higher energy state. If the collision in a half-period results in a drop of total mechanical energy, the particle can move into lower energy equilibrium. However, there is a critical value of energy increment which is balanced (damped) by the energy loss due to viscosity. When the energy increment accumulated by the particle is less than this critical value, the particle could not leave its original equilibrium position after a half-cycle. Only when the energy increment accumulated by the particle exceeds this critical value, could the particle migrate to its neighbor streamline and its equilibrium will become unstable.

The stability of a flow depends on the relative magnitude of $\Delta E$ and $\Delta H$. For flow with a curved streamline, with similar steps as in [11], the relative magnitude of the energy gained from collision and the energy loss due to viscous friction determines the disturbance amplification or decay. Thus, for a given flow, a stability criterion can be written as below for the half-period, by using Eq.(6) and Eq.(7),

$$F = \frac{\Delta E}{\Delta H} = \left(\frac{\partial E}{\partial n}\frac{2A}{\pi}\right) \bigg/ \left(\frac{\partial H}{\partial s}\frac{\pi}{\omega}u\right) = \frac{2}{\pi^2}K\frac{A\omega}{u} = \frac{2}{\pi^2}K\frac{v'_m}{u} < Const, \qquad (8)$$

and

$$K = \frac{\partial E/\partial n}{\partial H/\partial s}. \qquad (9)$$

Here, $F$ is a function of coordinates which expresses the ratio of the energy gained in a half-period by the particle and the energy loss due to viscosity in the half-period. $K$ is a dimensionless field variable (function) and expresses the ratio of transversal energy gradient and the rate of the energy loss along the streamline. Here, $E = p + \frac{1}{2}\rho V^2$ is the total mechanical energy, $s$ is along the streamwise direction and $n$ is along the transverse direction.

It can be found from Eq.(8) that the instability of a flow depends on the value of $K$ and the amplitude of the relative disturbance velocity $v'_m/u$. For given disturbance, the maximum of $K$, $K_{max}$, in the flow domain determines the stability. Therefore, $K_{max}$ is taken as a stability parameter here. For $K_{max}<K_c$, the flow is stable; for $K_{max}>K_c$, the flow is unstable. Here, $K_c$ is the critical value of $K_{max}$. For any type of flows, it can be demonstrated that the variable $K$ is proportional to the global Reynolds number for a given geometry [11,14]. Thus, it is found from Eq.(8) that the critical amplitude of the disturbance in curved flows scales with the Re by an exponent of -1, which is the same as in parallel flows.

**3. Derivations of theorems**

For pressure driven flows, the derivatives of the total mechanical energy in the two directions can be expressed, respectively, as [12][16],

$$\frac{\partial E}{\partial n} = \frac{\partial (p + (1/2)\rho u^2)}{\partial n} = \rho(\mathbf{u} \times \boldsymbol{\omega}) \cdot \frac{d\mathbf{n}}{|d\mathbf{n}|} + (\mu \nabla^2 \mathbf{u}) \cdot \frac{d\mathbf{n}}{|d\mathbf{n}|} = \rho u \omega + (\mu \nabla^2 \mathbf{u})_n, \quad (10)$$

$$\frac{\partial E}{\partial s} = \frac{\partial (p + (1/2)\rho u^2)}{\partial s} = \rho(\mathbf{u} \times \boldsymbol{\omega}) \cdot \frac{d\mathbf{s}}{|d\mathbf{s}|} + (\mu \nabla^2 \mathbf{u}) \cdot \frac{d\mathbf{s}}{|d\mathbf{s}|} = (\mu \nabla^2 \mathbf{u})_s, \quad (11)$$

where $\boldsymbol{\omega} = \nabla \times \mathbf{u}$ is the vorticity.

It is found from Eq.(10) that the transversal energy gradient is composed of two parts: the vorticity flux and the viscous diffusion of transversal velocity. For parallel flows, there is no viscous diffusion of transversal velocity. This is a significant difference between parallel and non-parallel flows, which may result in variations of flow phenomena between them. It should be noticed that the boundary layer flow on a flat plate is not a parallel flow, and therefore, there exists viscous diffusion of transversal velocity. This viscous diffusion would change the energy distribution and would cause more unstable role if it increases the value of K according to "energy gradient theory."

**Theorem (1): Potential flow (inviscid and $\nabla \times \mathbf{u} = \mathbf{0}$) is stable.**

**Proof:** For inviscid flow, there is no energy loss along the streamline due to without viscosity. From Eq.(11), we have

$$\frac{\partial H}{\partial s} = 0. \quad (12)$$

The energy gradient in the transverse direction for potential flow is [16],

$$\frac{\partial E}{\partial n} = \frac{\partial (p + 1/2\rho u^2)}{\partial n} = 0. \quad (13)$$

Introducing Eq.(12) and Eq.(13) into Eq.(9), the value of K is, everywhere,

$$K = \frac{\partial E / \partial n}{\partial H / \partial s} = \frac{0}{0}. \quad (14)$$

In this case, the value of K is indefinitive. We can do the following analysis. For potential flow, the mechanical energy is uniform in the flow field everywhere, the imposed disturbance could not be amplified without an energy gradient, no matter how large the disturbance amplitude is. As the result, we conclude that potential flow (inviscid and $\nabla \times \mathbf{u} = \mathbf{0}$) is stable. Therefore, *turbulence could not be generated in potential flows*. Uniform rectilinear flow is an example of potential flow in parallel flows. For the basic cases of potential flow such as uniform flow, source/sink, free vortex, and corner flow, they are always stable. This is able to explain why a tornado can sustain a long time and does not breakdown. This is also able to explain that a swirling flow in a radial vaneless diffuser between two parallel walls can sustain a stable flow and can get large pressure recovery for appropriate air angle from the circumferential direction.

**Theorem (2): Inviscid rotational ($\nabla \times \mathbf{u} \neq \mathbf{0}$) flow is unstable.**

**Proof:** For inviscid flow, there is no energy loss along the streamline due to without viscosity. From Eq.(11), we have

$$\frac{\partial H}{\partial s} = 0. \tag{15}$$

The energy gradient in the transverse direction (due to rotational) is [16], from Eq.(10),

$$\frac{\partial E}{\partial n} = \frac{\partial (p + 1/2 \rho u^2)}{\partial n} \neq 0. \tag{16}$$

Introducing Eq.(15) and Eq.(16) into Eq.(9), the value of K is,

$$K = \frac{\partial E / \partial n}{\partial H / \partial s} = \infty. \tag{17}$$

It is seen that for inviscid rotational flow, the transversal energy gradient is not zero, and the there is no viscous energy loss in streamline direction to damp the disturbance since it is an inviscid flow. Thus, any imposed finite disturbance could be amplified by the transversal energy gradient ($F = \infty$ in Eq.(8)) at enough high Re. Therefore, we conclude that inviscid rotational ($\nabla \times V \neq 0$) flow is unstable. This theorem has important significance for climate dynamics and meteorology, since most air flow over the atmosphere boundary layer can be treat as inviscid rotational. This mechanism may be dominating in the formation of tornado.

**Theorem (3): Velocity profile with an inflectional point is unstable when there is no work input or output to the system, for both inviscid and viscous flows.**

**Proof:** For inviscid flow, there is no energy loss along the streamline. For viscous flow, the energy loss due to viscosity is zero at the inflection point from Navier-Stokes equation ($\partial H/\partial s = |\partial E/\partial s| = |(\mu\nabla^2 \mathbf{V})_s| = 0$) if there is no work input or output to the system (meaning pressure driven flows) [12,17]. Thus, for both inviscid and viscous flows, we have along the streamline at the inflection point,

$$\frac{\partial H}{\partial s} = 0. \tag{18}$$

For inviscid flow, when there is an inflection point on the velocity profile ($\frac{\partial u}{\partial n} \neq 0$) and if it is not at the stationary wall ($u \neq 0$), the energy gradient in the transverse direction at this point generally (due to rotational) is

$$\frac{\partial E}{\partial n} = \frac{\partial(p + 1/2\rho u^2)}{\partial n} \neq 0. \tag{19}$$

The addition of viscosity only changes the distribution, and does not off its value to zero (Eq.(10)). Introducing Eq.(18) and Eq.(19) into Eq.(9), the energy gradient parameter K at this point is,

$$K = \frac{\partial E/\partial n}{\partial H/\partial s} = \infty. \tag{20}$$

Thus, the value of function *K* becomes infinite at the inflection point and indicates that the flow is unstable when it is subjected to a finite disturbance ($F = \infty$ in Eq.(8)). Therefore, we conclude that velocity profile with an inflectional point is unstable when there is no work input or output to the system for both viscous flow and inviscid flow. For both inviscid flow and viscous flow, this is a sufficient condition but not a necessary condition for instability. If there is work input or output to the system, $\partial H/\partial s \neq 0$ at the inflection point, then this theorem is not established anymore.

Velocity inflection could result in instability as found in experiments and simulations, e.g., the vortices behind the cylinder at an enough high Re. This phenomenon has been identified as inviscid instability in the community. According to present theory, no matter what the flow is inviscid or viscous, inflection necessarily leads to instability, not just inviscid instability. But, the fact is really true that viscosity does not play an important role when an inflection instability occurs since the energy loss due to viscosity is zero at the inflection point. However, for inviscid flow, that inflection leads to instability is also correct.

**4. Comparison with Experiments**

This theory has been applied to studies for Taylor-Couette flow between concentric rotating cylinders [18]. This theory obtains very good agreement with the available experimental data of Taylor-Couette flows in literature. For the occurrence of primary instability, the critical value of $K_{max}$ is a constant for a given geometry no matter how the rotating speeds of the two cylinders for all the available experiments. The critical value of $K_{max}$ is observed from the experiments at the condition of occurrence of primary instability for the case of the inner cylinder rotating and the outer cylinder set to rest. These results confirm that the proposed theory is also applicable to rotating flows.

In Taylor-Couette flow between concentric rotating cylinders, the streamlines are concentric cycles and the transversal velocity is zero. Thus, there is no viscous diffusion of transversal velocity. Therefore, viscosity has only stable role to the base flow which is similar to parallel flows. These are in agreement with the experimental observations.

**5. Rayleigh Theorem is incorrect**

Rayleigh's so-called point-of-inflection criterion states that the necessary condition for instability of inviscid flow is the existence of an inflection point on the velocity profile [1-6].

According to this criterion, if it is a necessary condition, then inviscid flow without an inflection point on the velocity profile is stable. This conclusion is contradicting to the Theorem 2 of this study. It is not difficult that the reader can judge which is correct between the two theorems. The Rayleigh's theorem is only deduced from mathematics and has no physical background. The reason why Rayleigh's theorem is incorrect can be found from the following.

(1) In the derivation of Rayleigh's theorem, two-dimensionality of disturbance is assumed. Then, the Rayleigh equation for inviscid flow is obtained using a stream function as a variable. As we know, the disturbance is necessarily three-dimensional in reality. An extension in one direction must cause the compression in the other two directions, and vice versa (Fig.2). Even if the given disturbance at the beginning is two-dimensional, it can develop into three-dimensional. As is well known, there is no stream function in three-dimensional flows. It is also remembered that there is no turbulence in two-dimensional flows [19]. Even if this issue for three-dimensionality of disturbance could be overcome via Squire's transformation that the three-dimensional disturbance is transformed into a pseudo two-dimensional disturbance, the following problem exists.

(2) Another issue is that the amplitude of the traveling wave disturbance is assumed to be irrelevant to the spanwise direction, and is only a function of transverse coordinate, $A=A(y)$. Actually, the amplitude of traveling wave disturbance may vary in spanwise direction with the its propagating, $A=A(y, z)$, and thus the maximum disturbance rotates with the axis in propagating direction. The amplitude of the resultant disturbance travels following a spiral trace in shear flows, and this has been demonstrated by numerical simulations and experiments [20,21,22]. Therefore, the assumption of $A=A(y)$ is incorrect and the obtained result does not accord with the physics of the flows.

If the above problems are taken into consideration, the derivation by Rayleigh is not established anymore. Thus, one is not able to deduce the Rayleigh theorem.

## 6. Conclusions

The "energy gradient theory" proposed in a previous study is applied to curved flows in this study. Then, three important theorems for flow instability are deduced. These theorems are of great significance for the understanding of flow phenomena and the explanation of complex flows. In parallel flows and Taylor-Couette flows, viscosity has only stable role to the flow owing to its leading to high energy loss which damps instability and there is no viscous diffusion caused by transversal velocity. In non-parallel flows, such as boundary layer flow, viscosity may have unstable role to the flow owing to that viscous diffusion of transversal velocity may change the transversal energy gradient, besides its stable role due to energy loss. Although at present we could not predict the critical Reynolds number and the relevant wave number for instability with the proposed "energy gradient theory," it is very useful to deduce these important theorems which help us to understand many complex flow phenomena in nature.

From the results in this study, it is concluded that the classical Rayleigh theorem on inflectional velocity instability of inviscid flows is incorrect which has last for more than a century. Therefore, Rayleigh theorem on inflectional instability should be corrected in future in the most text books.

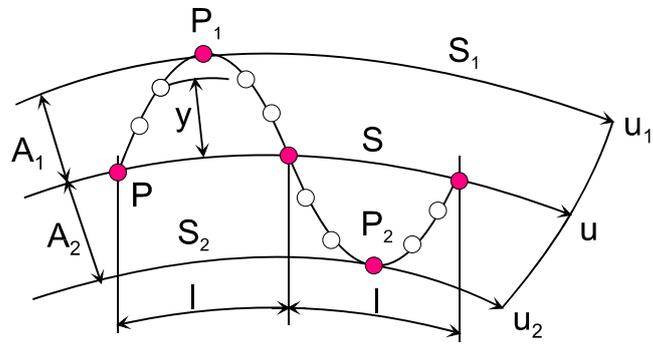

Fig.1 Movement of a particle around its original equilibrium position in a cycle of disturbance for curved flows.

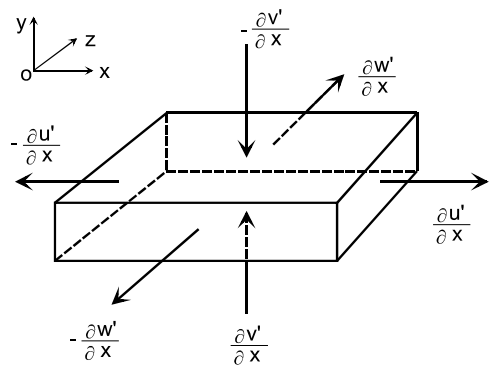

Fig.2 Compression in y direction leading to elongation x and z directions.

13